\documentclass[aps,prb,twocolumn,longbibliography,superscriptaddress,article]{revtex4-2}
\usepackage{epsfig}
\usepackage{epstopdf}
\usepackage{amsmath}
\usepackage{amsfonts}
\usepackage{amssymb}
\usepackage{hyperref}
\usepackage{bm}
\usepackage{makecell}
\usepackage{rotating}%
\usepackage{hyperref}
\usepackage{multirow}%
\usepackage{graphicx}
\usepackage{float}

\usepackage{graphicx}
\usepackage{dcolumn}
\usepackage{bm}
\usepackage{color}

\usepackage{tikz,xcolor,hyperref}

\definecolor{lime}{HTML}{A6CE39}
\DeclareRobustCommand{\orcidicon}{%
	\begin{tikzpicture}
	\draw[lime, fill=lime] (0,0)
	circle [radius=0.16]
	node[white] {{\fontfamily{qag}\selectfont \tiny ID}};
	\draw[white, fill=white] (-0.0625,0.095)
	circle [radius=0.007];
	\end{tikzpicture}
	\hspace{-2mm}
}

\foreach \x in {A, ..., Z}{%
	\expandafter\xdef\csname orcid\x\endcsname{\noexpand\href{https://orcid.org/\csname orcidauthor\x\endcsname}{\noexpand\orcidicon}}
}

\begin{document}

\title{Magnetically tunable symmetry-enforced nodal lines producing \\ huge anomalous Hall conductivity in altermagnetic $\alpha$-MnTe}

\author{Mathews Benny\orcidM}
\affiliation{International Research Centre Magtop, Institute of Physics, Polish Academy of Sciences, Aleja Lotnik\'ow 32/46, PL-02668 Warsaw, Poland}

\author{Xujia Gong\orcidA}
\affiliation{International Research Centre Magtop, Institute of Physics, Polish Academy of Sciences, Aleja Lotnik\'ow 32/46, PL-02668 Warsaw, Poland}

\author{Amar Fakhredine\orcidF}
\email{amarf@ifpan.edu.pl}
\affiliation{Institute of Physics, Polish Academy of Sciences, Aleja Lotnik\'ow 32/46, 02668 Warsaw, Poland}

\author{Raphaël Salazar\orcidR}
\affiliation{New Technologies-Research Center, University of West Bohemia in Pilsen, Teslova 9a, 301 00, Pilsen, Czech Republic}

\author{Ashutosh S. Wadge\orcidW}
\affiliation{International Research Centre Magtop, Institute of Physics, Polish Academy of Sciences,
Aleja Lotnik\'ow 32/46, PL-02668 Warsaw, Poland}

\author{Juraj Krempaský} 
\affiliation{Photon Science Division, Paul Scherrer Institut, CH-5232 Villigen, Switzerland}

\author{Gunther Springholz}  
\affiliation{Institute of Semiconductor and Solid State Physics, Johannes Kepler University of Linz, Linz, Austria}

\author{Sarath Sasi\orcidX} 
\affiliation{New Technologies-Research Center, University of West Bohemia in Pilsen, Teslova 9a, 301 00, Pilsen, Czech Republic}

\author{Mahdi Hajloui}  
\affiliation{Institute of Semiconductor and Solid State Physics, Johannes Kepler University of Linz, Linz, Austria}

\author{Martin Heinrich\orcidH}  
\affiliation{Institute of Semiconductor and Solid State Physics, Johannes Kepler University of Linz, Linz, Austria}

\author{Dawid Wutke}
\affiliation{Solaris National Synchrotron Radiation Centre, Jagiellonian University, Czerwone Maki 98, 30-392 Kraków, Poland}

\author{Rafał Kurleto}
\affiliation{Solaris National Synchrotron Radiation Centre, Jagiellonian University, Czerwone Maki 98, 30-392 Kraków, Poland}
 
\author{Natalia Olszowska\orcidN}
\affiliation{Solaris National Synchrotron Radiation Centre, Jagiellonian University, Czerwone Maki 98, 30-392 Kraków, Poland}

\author{Sahar Izadi Vishkayi\orcidI}
\affiliation{International Research Centre Magtop, Institute of Physics, Polish Academy of Sciences, Aleja Lotnik\'ow 32/46, PL-02668 Warsaw, Poland}

\author{Asiyeh Shokri\orcidG}
\affiliation{International Research Centre Magtop, Institute of Physics, Polish Academy of Sciences,
Aleja Lotnik\'ow 32/46, PL-02668 Warsaw, Poland}

\author{Ján Minár } 
\affiliation{New Technologies-Research Center, University of West Bohemia in Pilsen, Teslova 9a, 301 00, Pilsen, Czech Republic}

\author{Carmine Ortix\orcidO}
\affiliation{Dipartimento di Fisica “E. R. Caianiello”, Universit\'a di Salerno, IT-84084 Fisciano (SA), Italy}

\author{Jeroen van den Brink\orcidJ}
\affiliation{Leibniz-Institut f\"ur Festk\"orper- und Werkstoffforschung (IFW Dresden), Helmholtzstr. 20, D-01069 Dresden, Germany}

\author{Jakub Schusser\orcidS}
\email{schusser@ntc.zcu.cz}
\affiliation{New Technologies-Research Center, University of West Bohemia in Pilsen, Teslova 9a, 301 00, Pilsen, Czech Republic}

\author{Carmine Autieri\orcidB}
\email{autieri@magtop.ifpan.edu.pl}
\affiliation{International Research Centre Magtop, Institute of Physics, Polish Academy of Sciences,
Aleja Lotnik\'ow 32/46, PL-02668 Warsaw, Poland}

\begin{abstract}
Altermagnetic $\alpha$-MnTe exhibits huge anomalous Hall conductivity (AHC) up to room-temperature with weak ferromagnetism arising from spin and orbital polarizations. We clarify the origin of the large value of the AHC by identifying two sets of distinct symmetry-enforced nodal lines in the valence bands, located at $k_z=0$ and $k_z=\frac{\pi}{c}$, protected by mirror symmetry $M_z$ and glide symmetry $G_z = \{M_z\,|\,0,0,\tfrac{c}{2}\}$, respectively. Both nodal lines are energy-dependent with an approximate C$_6$ symmetry, which is reduced to an exact C$_2$ symmetry due to the presence of the N\'eel vector.
The highest valence band exhibits a Mexican-hat dispersion, whereas the second-highest valence band exhibits an inverted Mexican-hat dispersion, with nodal lines at the crossing between them. 
Within first-principles accuracy, we demonstrate that these nodal lines give rise to the large AHC observed experimentally and exhibit a strong interplay with the weak ferromagnetism. We further show that even a small spin canting strongly modifies the nodal lines and the AHC, making them both magnetically tunable. By disentangling the altermagnetic and ferromagnetic contributions to the AHC, the altermagnetic contribution dominates at small canting angles, while the ferromagnetic contribution becomes sizeable for larger values. Using linear dichroism in angle-resolved photoemission spectroscopy, we show a signature of the nodal line at the border of the Brillouin zone. 
\end{abstract}


\maketitle

\noindent\textbf{Introduction:}
In altermagnets, sites with opposite spins are connected by rotational symmetries, either proper or improper, and symmorphic or nonsymmorphic, but not by translation or inversion symmetries\cite{Smejkal22beyond,doi:10.1126/sciadv.aaz8809,hayami2019momentum,hayami2020bottom,Smejkal22,yuan2023degeneracy,Samanta2025,Sun2025,D3NR03681B,D3NR04798A,ssxp-gz9l,D4NR04053H,https://doi.org/10.1002/adfm.202505145,Berritta2025,jr65-4273,ref1,Ma2021,sorn2026projectedaltermagnetismsymmetryreduction,plucinski2026multipletselectivephotoelectrondiffractionaltermagnet,dou2026electricfieldswitchablemagneticspin,10.1088/1361-6633/ae8868,Fakhredine25b}. While the breaking of time-reversal symmetry and the weak ferromagnetism induced by spin-orbit coupling (SOC) was known\cite{DZYALOSHINSKY1958241}, one of the most striking features of altermagnets is the non-relativistic spin–momentum locking with even-wave symmetry. Non-relativistic spin–momentum locking refers to a phenomenon in which the electron spin orientation is locked to its crystal momentum. This locking ensures a symmetry-protected zero net magnetization in the non-relativistic limit.
Some classes of altermagnets can exhibit anomalous Hall conductivity (AHC), depending on the magnetic space group, microscopic details, and the orientation of the Néel vector; which is defined as the difference between the spin vectors at the two magnetic sites \cite{839n-rckn,PhysRevB.111.184407}. The simplest form of antisymmetric exchange interaction generating the AHC is the staggered Dzyaloshinskii–Moriya interaction\cite{DZYALOSHINSKY1958241}, which produces relativistic weak ferromagnetism orthogonal to the Néel vector\cite{PhysRevB.111.054442}. When SOC is included, the non-relativistic spin–momentum locking is inherited by the dominant spin component\cite{Fakhredine26}, while the other two components are referred to as subdominant and tend to exhibit $d$-wave spin–momentum locking as well. We refer to this effect across all spin components as the relativistic spin–momentum locking \cite{Fakhredine26,PhysRevB.109.024404,PhysRevB.110.144412}.

MnTe crystallizes in space group P6$_3$/mmc (no. 194), adopting the NiAs crystal structure, and exhibits mirror symmetry $M_z$ as well as nonsymmorphic symmetry\cite{Takahashi2025}. MnTe is a nonrelativistic $g$-wave altermagnet\cite{Krempask2024,Amin2024,PhysRevB.109.115102,k36v-91br,uykur2026revisitingsymmetryopticalphonons,32gt-3nfn,dzk6-f68q,doi:10.1021/acs.nanolett.5c01158,weber2026magneticfieldcontrolneel} while it exhibits a $d$-wave character in the relativistic spin-momentum locking\cite{Fakhredine26} allowing noncollinear spin current\cite{Song2026}. Correspondingly, the spin splitting is absent along four nodal planes in the non-relativistic limit, while in the relativistic regime it is everywhere except at $\Gamma$ \cite{Fakhredine26,gong2026symmetryprotectednodalplanesaccidental}.
MnTe also exhibits an AHC\cite{PhysRevLett.130.036702,chen2026strainengineeringintrinsicanomalous,sarkar2026anomaloushalleffectsiliconcompatible,nyy3-h8wr,bey2026conductivityscalinganomaloushall}, arising when the easy axis is along the $y$-direction\cite{PhysRevB.96.214418,bb6r-2nwz} with a net magnetization\cite{kluczyk2023coexistence} dominated by the orbital magnetic moment\cite{g32j-hnvz, ren2026atomicscaleobservationsymmetrybreaking,sandratskii2026spinorbitcouplingeffectsaltermagnets,zhao2026residualorbitalmagnetizationgoverns}. 
The spin canting is governed by a higher-order staggered Dzyaloshinskii–Moriya interaction.\cite{PhysRevB.111.054442,g32j-hnvz}. 
Beyond this weak ferromagnetism, the $S_z$ component also exhibits d-wave spin–momentum locking of the $Q_{x^2-y^2}$ wave\cite{hirakida2025multipoleanalysisspincurrents, usanov2026discerninggroundstatephotoemissioninduced,din2025unconventionalrelativisticspinpolarization}. Its altermagnetic order can be switched\cite{yklc-9n6t} and be used as an efficient source and detector of spin current\cite{PhysRevLett.134.086701}.

\begin{figure}
    \centering
    \includegraphics[width=1\linewidth]{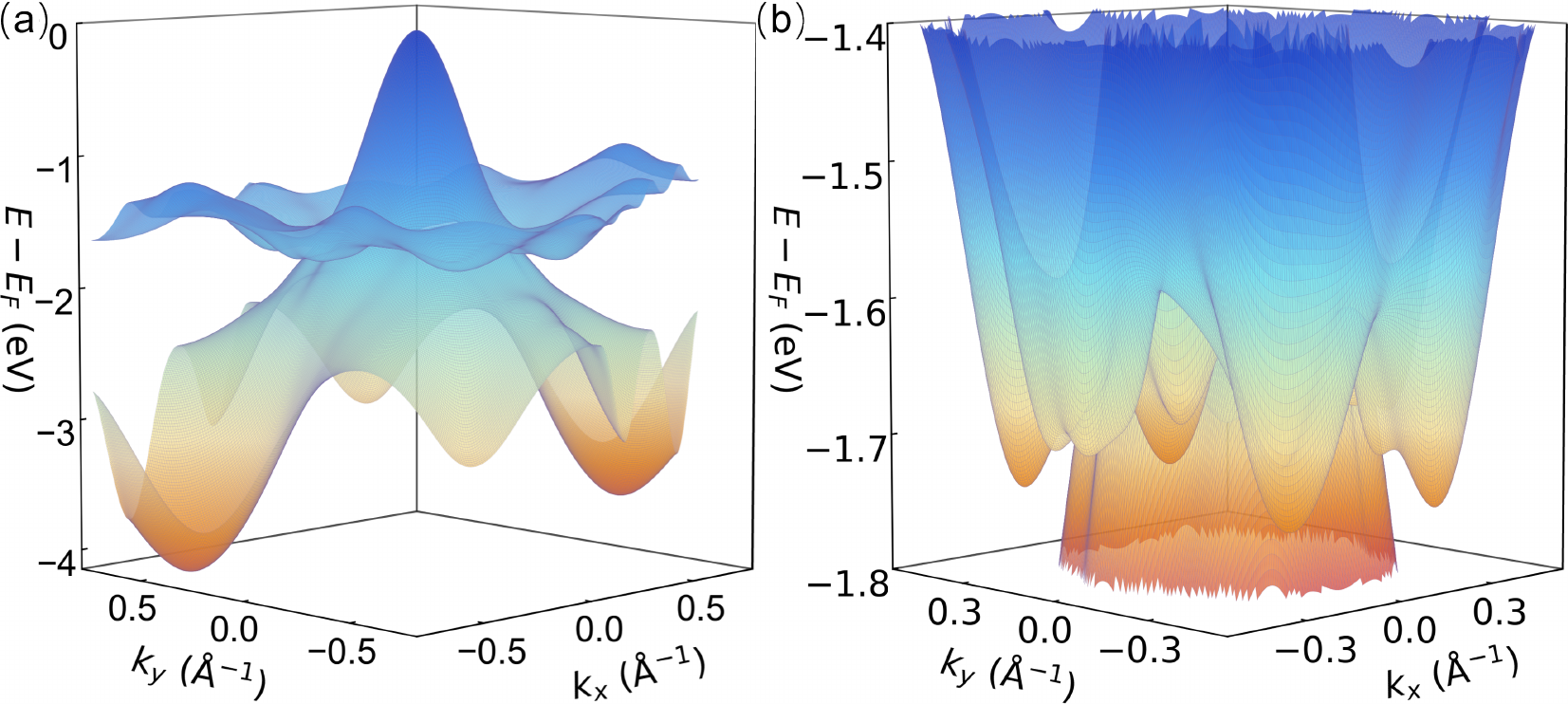}
    \caption{Band dispersion of the valence band of $\alpha$-MnTe in the $k_z=\frac{\pi}{c}$ plane. (a) Energy range from $-4.0$ to $0$ eV. (b) Magnified view of the energy range from $-1.8$ to $-1.4$ eV. The Coulomb repulsion parameter is $U=4$ eV. The surface height and color both represent $E-E_F$ (in eV), with red indicating lower energies and blue indicating higher energies.}
    \label{fig1:3dfermisurface_u4}  
\end{figure}


Topological nodal-line semimetals are characterized by band crossings that form one-dimensional loops in momentum space~\cite{wadge2026contemporaryinsightselectronicstructure,Muhammad2024}, giving rise to nontrivial Berry phases and enhanced Berry curvature effects, which were measured in non-magnetic systems\cite{Veyrat2025-jk}, Kramers degenerate antiferromagnets\cite{Hosen2018-do}, ferromagnets\cite{Kim2018,Clark2026-ev,Sadhukhan2025}, ferrimagnets\cite{Li2025-hw} and parity-mixed antiferromagnets\cite{terada2026nodallineenhancedquantumgeometriceffects}. The stability of these nodal lines is dictated by crystalline symmetries that prevent hybridization between crossing bands. Both mirror and glide mirror symmetries can enforce degeneracies in mirror-invariant planes, for instance at k$_z$=0,$\frac{\pi}{c}$, when the intersecting bands carry opposite mirror eigenvalues, ensuring topological protection.
Notable examples are the non-magnetic non-centrosymmetric compound PtBi$_2$\cite{Bian2016-kw,Veyrat2025-jk} and IV–VI semiconductors\cite{PhysRevLett.122.186801}.

\begin{figure*}[t]
    \centering
    \includegraphics[width=0.99\textwidth]{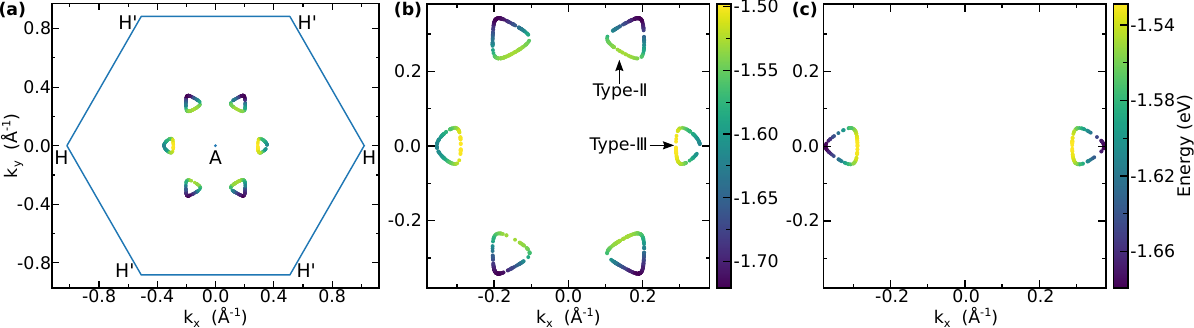}
    \caption{Nodal line at $k_z=\frac{\pi}{c}$ without canting (a) plotted within the first BZ and (b) its magnification. The border of the BZ is marked in blue. (c) Nodal line at $k_z=\frac{\pi}{c}$ with a canting of $10^\circ$ where only the type-III nodal line survives.}
    \label{fig2:nodal_line_k05_canting} 
\end{figure*}

CrSb exhibits the same space group as MnTe but with a different N\'eel vector direction, and it has been demonstrated to host symmetry-protected Weyl points in its bulk electronic structure. CrSb exhibits a nodal line in the non-relativistic case, which evolves into Weyl points upon including SOC with topological Fermi-arcs\cite{Li2025}. These Weyl points generate strong Berry curvature hotspots; however, the net AHC cancels out due to symmetry constraints. The nodal line was also reported in the rutile material class\cite{PhysRevB.95.235104} and in the pure altermagnet V$_2$Te$_2$O\cite{hu2026observationspinvalleylockednodal}.
In $\alpha$-MnTe, we identify two symmetry-distinct nodal lines: a mirror-protected nodal line at $k_z=0$ and a non-symmorphic-symmetry-protected nodal line at $k_z=\frac{\pi}{c}$. Unlike CrSb, RuO$_2$, and V$_2$Te$_2$O, $\alpha$-MnTe exhibits an AHC, making it an ideal platform for the systematic study of the interplay among nodal lines, weak ferromagnetism, and the AHC. These nodal structures are expected to act as sources of Berry curvature, thus becoming relevant for the AHC. We investigate the sensitivity of the Berry curvature distribution and the AHC stemming from the nodal line to a finite canting angle. While the band structure is weakly modified by a small canting, the Berry curvature is highly responsive, leading to pronounced changes in the AHC.


\noindent\textbf{Density functional theory and relativistic Hamiltonian within ab-initio accuracy:}
We perform density functional theory (DFT) calculations for the altermagnetic phase. Fig.~\ref{fig1:3dfermisurface_u4}(a) shows the energy as a function of $k_x$ and $k_y$ at $k_z=\frac{\pi}{c}$. The point $(k_x,k_y)=(0,0)$ corresponds to the A point, where the top valence band reaches a maximum and disperses downward with a hole-like effective mass to approximately $-1.5$~eV, where it exhibits a local minimum before dispersing upward again. Therefore, the highest valence band at $k_z=\frac{\pi}{c}$ forms a Mexican hat, while the second-highest valence band displays an inverted Mexican hat. A magnified view in Fig.~\ref{fig1:3dfermisurface_u4}(b) highlights the region relevant to the nodal line, which lies between $-1.7$ and $-1.5$~eV. Additional band-structure calculations for different values of the Coulomb repulsion are provided in the Supplementary Materials, demonstrating the robustness of the nodal line. Experimental samples of $\alpha$-MnTe are intrinsically $p$-doped; therefore, the presence of the nodal line in the valence band strongly affects the transport.

To study the interplay among the nodal line, weak ferromagnetism, and AHC, we adopt a strategy\cite{benny2026staggereddzyaloshinskiimoriyacantingangle,jskol_SOC_Code_V1_2025} developed recently to extract the nonmagnetic Hamiltonian from the Wannierization procedure and systematically tune the magnetic moments, canting angles, and SOC described in the Supplementary Materials. This ensures the correct magnetic and crystal symmetries of the Hamiltonian and, consequently, the correct symmetries of the AHC.
According to the magnetic symmetry of MnTe, the spin can be described by the vectors
$\mathbf{S_{\mathrm{Mn1}}}$ = $(0, +S\cos\theta, S\sin\theta)$
and 
$\mathbf{S_{\mathrm{Mn2}}}$=$(0, -S\cos\theta, S\sin\theta)$
with the N\'eel vector along the $y$-axis and spin canting along the $z$-axis. 
These equations define $\theta$ as the canting angle of the Mn spins\cite{PhysRevB.111.054442, g32j-hnvz, Fakhredine26}. The experimental value is expected to lie in the range $\theta$=0.0001-0.03, depending on the estimation of weak ferromagnetism\cite{kluczyk2023coexistence,g32j-hnvz,ren2026atomicscaleobservationsymmetrybreaking}.

\noindent\textbf{Nodal lines and their interplay with weak ferromagnetism:}
we analyze the band structure to identify nodal lines through a model Hamiltonian with a tunable canting angle. We find two distinct nodal lines located at $k_z = 0$ and $k_z = \frac{\pi}{c}$, both robust across the relevant range of model parameters. At $k_z=0$ and $k_z=\frac{\pi}{c}$, the spin components $S_x$ and $S_y$ vanish\cite{Fakhredine26}, as these planes correspond to nodal planes for their relativistic spin-momentum lockings. Therefore, only the $S_z$ component survives, with the spin-momentum locking characterized by the Q$_{x^2-y^2}$ symmetry dictating the structure of the nodal line. As a consequence, the k$_x$=0 and k$_y$=0 are mirror planes of the system.

\begin{figure}
    \centering
    \includegraphics[width=0.99\linewidth,angle=0]{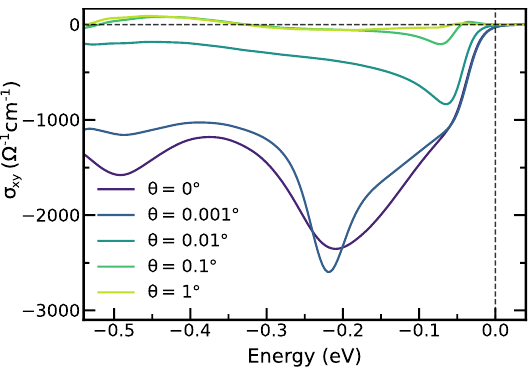}
    \caption{$\sigma_{xy}$ as a function of the energy for different canting angles $\theta$, focusing on the energies near the top valence band.}
    \label{fig:canting_angle}
\end{figure}

We denote the set of nodal lines at $k_z = \frac{\pi}{c}$ as NL1 and the set of nodal lines at $k_z = 0$ as NL2.
The collinear magnetic ordering gives rise only to NL1, consistent with DFT calculations. NL1 consists of six warped triangles arranged with an approximately C$_6$-like symmetry as reported in Fig.~\ref{fig2:nodal_line_k05_canting}(a); however, in MnTe, the exact symmetry is C$_2$, while the apparent C$_6$ symmetry emerges only in the non-relativistic limit. The warped triangles are located at intermediate wave vectors surrounding the A point at the center of the Brillouin zone (BZ) face. This nodal line is energy-dependent and lies between the highest and the second-highest valence band. 
More in detail, NL1 is composed of two subgroups of nodal lines, as we can see from Fig.~\ref{fig2:nodal_line_k05_canting}(b). Two equivalent warped triangles appear along the direction A-H, while the other four warped triangles appear along the direction A-H'. The inequivalence between the two subgroups arises due to the breaking of the C$_6$ symmetry and is further strengthened by the Q$_{x^2-y^2}$ spin-momentum locking of S$_z$\cite{g32j-hnvz}. The four warped triangles are type-II nodal lines, with the two crossing bands having the same sign of velocity, while the two warped triangles are type-III nodal lines, with the two bands sharing the same sign of velocity along one momentum direction but exhibiting opposite signs along the orthogonal direction.
By tuning the canting angle, one subgroup of the nodal line at $k_z=\frac{\pi}{c}$ persists up to approximately $10^\circ$, as we can see from Fig.~\ref{fig2:nodal_line_k05_canting}(c), while the other subgroup can persist up to approximately $20^\circ$. 
NL2 with C$_2$ symmetry is also energy-dependent and lies in the energy range between -1.05 and -1.45 eV. One subgroup of the nodal lines emerges for canting angles of $1.5^\circ$, which is why NL2 is not present in DFT calculations. A second subgroup of NL2 appears for $\theta$=20$^\circ$. 
These nodal lines both arise in the valence bands and are protected by mirror symmetry $M_z$ and glide symmetry $G_z = \{M_z\,|\,0,0,c/2\}$, respectively. 
To validate the role played by the glide symmetry in stabilizing NL1, we induced a canting for only one Mn sublattice. It was observed that such asymmetric canting of the order of 0.1$^\circ$ was sufficient to eliminate this nodal line.

\begin{figure*}[t]
    \centering
    \includegraphics[width=0.99\textwidth]{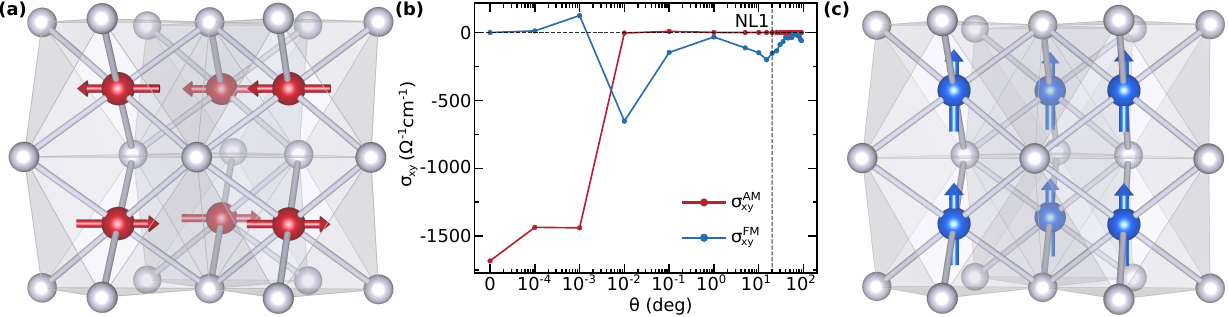}
    \caption{Disentangling the AHC of (a) the altermagnetic phase and (c) the ferromagnetic phase. (b) The AHC is plotted as a function of the canting angle on a logarithmic scale at an energy of 100 meV below the top valence band; the dashed vertical line represents the canting angle at which NL1 gaps out.}
    \label{fig:Figure5} 
\end{figure*}

We calculate the AHC of the altermagnetic phase with the N\'eel vector oriented along the $y$-axis. The canting angle in $\alpha$-MnTe is suppressed\cite{g32j-hnvz}; however, it increases at the surface, especially since the N\'eel vector is orthogonal to the surface \cite{BennyGong27,PhysRevX.14.021033,zhou2026surfacestatedrivenanomaloushalleffect,zhao2026emergentanomaloushalleffect}. 
The AHC as a function of the canting angle is reported in Fig.~\ref{fig:canting_angle}. A peak at 1.6 eV below the top valence band is associated with NL1, as confirmed by our analysis of the Berry curvature. There is also another peak located at 0.9~eV below the Fermi level, which is not generated at $k_z=\frac{\pi}{c}$ but arises anyway from the anticrossing between the same bands that form the Mexican-hat dispersion at $k_z=\frac{\pi}{c}$. The largest contribution to the AHC is provided by NL1, whereas the contribution from the $k_z=0$ plane is negligible, although it increases as soon as NL2 emerges.

\noindent\textbf{Disentangling ferromagnetic and altermagnetic contribution to the AHC:}
When the canting angle is zero, the system is a collinear altermagnet represented in Fig.~\ref{fig:Figure5}(a), whereas when the canting angle is 90$^\circ$, it is a collinear ferromagnet represented in Fig.~\ref{fig:Figure5}(c) with a continuous transformation from the altermagnetic to the ferromagnetic phase. 
To gain insight into the AHC, we decouple the ferromagnetic collinear and the altermagnetic collinear contributions.
To disentangle the total AHC into the ferromagnetic $\sigma_{xy}^{\mathrm{FM}}$ and altermagnetic $\sigma_{xy}^{\mathrm{AM}}$ parts, we use\cite{doi:10.1126/sciadv.aaz8809}:
\begingroup
\setlength{\jot}{1pt} 
\begin{equation}
\begin{aligned}
\sigma_{xy}^{\mathrm{AM}}
&= \frac{\sigma_{xy}(\mathbf{N},\mathbf{M}) + \sigma_{xy}(\mathbf{N},-\mathbf{M})}{2},\\
\sigma_{xy}^{\mathrm{FM}}
&= \frac{\sigma_{xy}(\mathbf{N},\mathbf{M}) - \sigma_{xy}(\mathbf{N},-\mathbf{M})}{2}.
\end{aligned}
\end{equation}
\endgroup
with $\mathbf{M}$ as the net magnetization and $\mathbf{N}$ as the N\'eel vector.
While previous studies achieved the disentanglement procedure only for rutile structures\cite{doi:10.1126/sciadv.aaz8809,bu2026disentanglinganomaloushalleffect} and model Hamiltonians\cite{Kipp2021}, the method developed here enables numerically stable calculations for magnetic systems with DFT accuracy. The technical details of the implementation are provided in the accompanying code documentation \cite{jskol_SOC_Code_V1_2025}. The results of disentangling the AHC for MnTe are reported in Fig.~\ref{fig:Figure5}(b) for the value of 0.1 eV below the maximum of the valence band, which is the experimental value of the Fermi level\cite{kluczyk2023coexistence}. 
The AHC is dominated by the contribution at k$_z$=$\frac{\pi}{c}$ and this contribution collapses when NL1 gaps out. Therefore, only NL1 is relevant for the AHC.
We observe that $\sigma_{xy}^{\mathrm{AM}}$ dominates at the angles $10^{-3}$-$10^{-4}$ as reported in Fig.~\ref{fig:Figure5}(b). At larger canting angles, the altermagnetic contribution becomes small and the ferromagnetic AHC is the major contributor. More generally, we can conclude that in an altermagnet, $\sigma_{xy}^{\mathrm{AM}}$ is dominant, but $\sigma_{xy}^{\mathrm{FM}}$ can be sizable and, occasionally, larger than the altermagnetic AHC (see Supplementary Materials). The AHC decreases with increasing net magnetization as a result of the disappearance of NL1, which is responsible for the large AHC.

\begin{figure*}
    \centering
    \includegraphics[width=0.95\linewidth,angle=0]{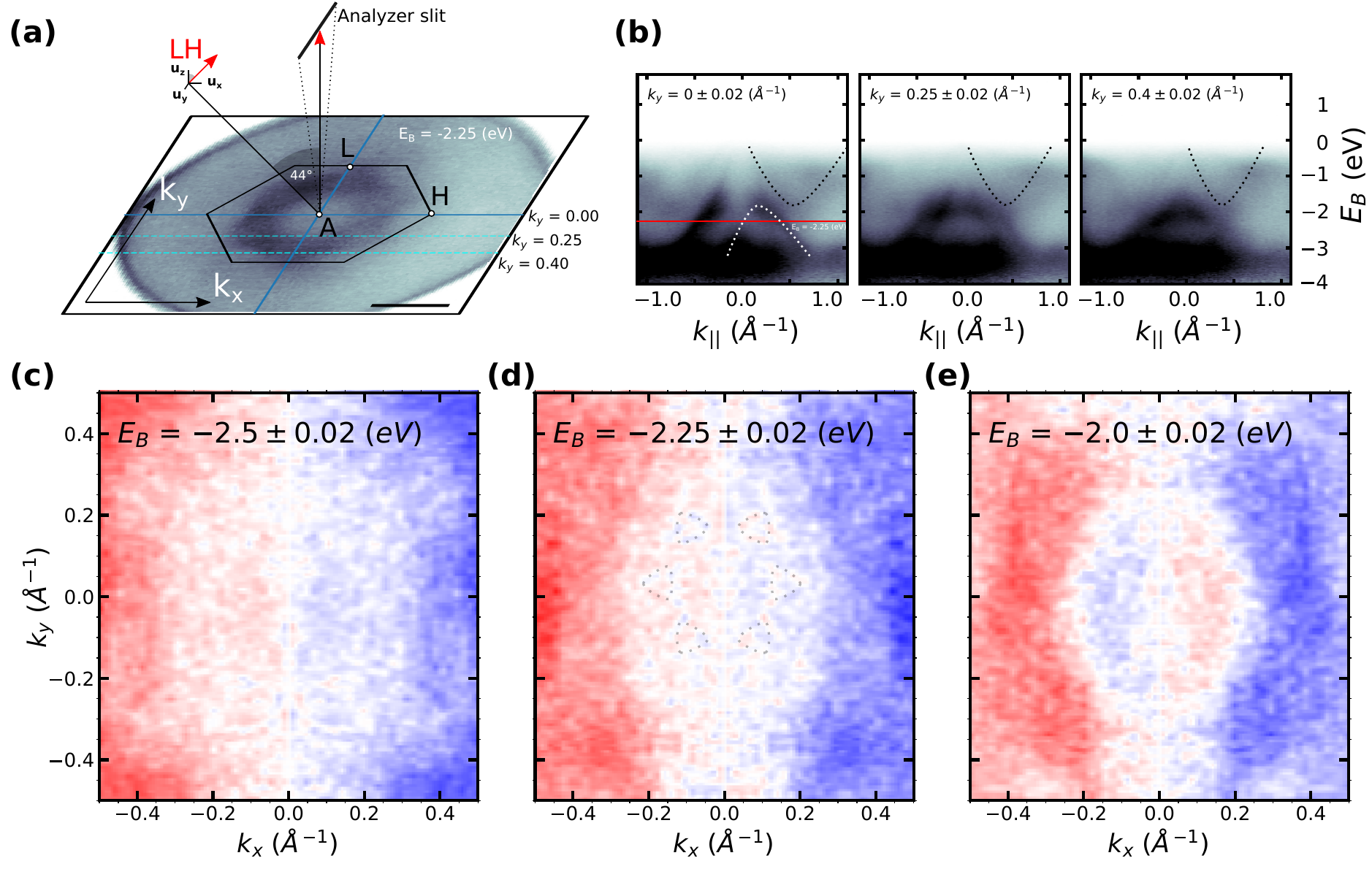}
    \caption{Experimental signatures of NL1: (a) Experimental geometry; the thick black line indicates the scale {0.5} $\mathrm{\AA}^{-1}$. The $p$-polarized (LH) light arrives at the sample with the detector slit aligned along the A-H direction. (b) Corresponding ARPES energy-momentum cuts indicating the overlap point of the Mexican-hat bands in the middle panel. (c)-(e) Linear dichroism constant-energy cuts. The dichroic plot in (d) captures abrupt dichroism sign change at the predicted nodal line positions.}
    \label{figX:Experiment}
\end{figure*}

\noindent\textbf{Experimental results:}
Linear~\cite{Schusser2022,Figgemeier2025}and circular dichroism~\cite{Unzelmann2021,Figgemeier2025} in angle-resolved photoemission spectroscopy (ARPES) were shown to be powerful probes of the orbital texture of electronic states, owing to their sensitivity to orbital angular momentum (OAM)~\cite{Schusser2022,Figgemeier2025,Erhardt2024,Unzelmann2021}, whose momentum-space texture correlates with Berry curvature~\cite{Schuler2020,Park2012,Park2012a,Oh2025,Cho2018} in several material classes. 
In Weyl semimetals, nodal-line semimetals, and other topological materials, abrupt sign changes of the dichroic signal have been observed across topological band crossings~\cite{Schusser2022,Figgemeier2025,Unzelmann2021,Oh2025,Sidilkover2025}.
Motivated by the methodology discussed in~\cite{Figgemeier2025}, we turn our attention to dichroic ARPES experiments on MnTe at the calculated position of NL1. We extract linear dichroism (LD) as defined in the Supplementary Materials, measured in the geometry indicated in Fig.~\ref{figX:Experiment} (a).
In Fig.~\ref{figX:Experiment}(c-e), we observe abrupt modulations of the photoemission intensity at the expected in-plane momentum values that move concentrically towards the $A$ point as a function of increasing binding energy, providing spectroscopic evidence consistent with the occurrence of nodal lines. Even though the proportionality of the dichroic signal in non-centrosymmetric systems to OAM is valid only along the high-symmetry direction, the LD modulations we observe are surprisingly reminiscent of the nodal line warped-triangle shape. Limited by the experimental resolution, we refrain from directly denoting the features reminiscent of six warped triangles as nodal lines. Instead, we interpret them as spectroscopic signatures occurring at the predicted position of NL1.

These experimental data were obtained on multi-domain samples with no preferential N\'eel vector orientation, suggesting preserved 6-fold symmetry of the observed features. It should be noted that the OAM and concomitant dichroic signal sign inversion in systems such as TaAs~\cite{Unzelmann2021,Figgemeier2025} stem from the non-centrosymmetry of the unit cell. In contrast, in altermagnetic MnTe, the Parity symmetry is retained, so the bulk inversion symmetry breaking mechanism established in TaAs cannot directly account for the sign change in the observed LD. Yet the observation of six warped triangles, at which the LD exhibits a clear sign reversal, remarkably matches the positions of NL1. With conserved parity symmetry, a plausible explanation of why we observe these features is locally broken inversion symmetry and sensitivity to local (hidden) orbital polarization, unraveled by the surface sensitivity of ARPES measurements that sample the two Te sublattices with different weights and provide a local OAM probe~\cite{Sidilkover2025}, analogous to hidden orbital texture~\cite{Schusser2024}. Importantly, we would like to comment on the position of the $E_{F}$. Unlike previous bulk-sensitive studies, the enhanced surface sensitivity of the vacuum ultraviolet measurements (h$\nu$=94.5 eV, A point) reveals an additional surface-derived state that lies a few hundred meV above the energy previously reported as the Fermi level~\cite{Krempask2024,Hajlaoui2024}. Incorporating this state into energy calibration leads to a revised value of E$_F$ (more details on E$_F$ in the Supplementary Materials). Despite the corresponding shift of the E$_B$ values, the agreement with bulk calculations remains unaffected. As indicated in the bare intensity spectra in Fig.~\ref{figX:Experiment}(b), the dichroic signal was extracted at the overlap of the Mexican-hat bands. 

\noindent\textbf{Conclusions and Outlook:} We have demonstrated the existence of nodal lines in $\alpha$-MnTe. The nodal line and the consequent large Berry curvature can be tuned magnetically through the spin canting angle. This tunability originates from the strong coupling between the canting angle and the topological electronic structure, with two distinct symmetry-enforced nodal lines playing a central role. NL1, located at $k_z = \frac{\pi}{c}$ is protected by the glide symmetry $G_z = \{ M_z \,|\, 0,0,c/2 \}$, while NL2 at $k_z = 0$, is protected by the mirror symmetry $M_z$. The symmetry-enforced nodal line at k$_z$=$\frac{\pi}{c}$ was observed experimentally. This nodal line is composed of two inequivalent subgroups: one exhibits type-II nodal-line behavior, while the other exhibits type-III nodal-line behavior.
The nodal lines follow the d-wave symmetry $Q_{x^2-y^2}$ of the S$_z$ component and drive the large AHC observed in MnTe by enhancing the Berry curvature. The altermagnetic AHC component dominates at small canting angles, while the ferromagnetic AHC contribution remains substantial. Finally, using LD in ARPES, we have shown the experimental signature of the nodal line. These results can open a research field on room-temperature nodal lines in altermagnets with weak ferromagnetism.

The presence of nodal lines in MnTe calls for a check of nodal lines also in the isostructural h-FeS\cite{yao2026uniaxialstraintunedmagnetism}, as well as for ferromagnets with the same space group as CrTe\cite{gong2026relativisticspinmomentumlockingferromagnets} and MnBi\cite{10.1063/5.0135578,11596489}, where the nodal lines could be closer to the Fermi level due to different fillings of the d-manifolds. These symmetry-protected nodal lines are expected to be tunable via strain or external pressure, enabling controllable topological responses~\cite{yao2026uniaxialstraintunedmagnetism, giri2026straintopologicalselectoraltermagnetic}. Our work establishes symmetry-enforced nodal lines as the microscopic mechanism underlying the anomalous Hall effect of $\alpha$-MnTe and identifying spin canting as a practical knob for controlling topological transport in altermagnets.

\section*{Acknowledgments}
The authors thank L. \v{S}mejkal, M. Gryglas-Borysiewicz, M. Bia{\l}ek and J. Skolimowski for useful discussions.
C. A. was supported by the Polish National Agency for Academic Exchange (NAWA) under the Bekker Programme, grant no. BPN/BEK/2025/1/00244/DEC/1.
This research was supported by the "MagTop" project (FENG.02.01-IP.05-0028/23) carried out within the "International Research Agendas" programme of the Foundation for Polish Science, co-financed by the European Union under the European Funds for Smart Economy 2021-2027 (FENG). We further acknowledge access to the computing facilities of the Interdisciplinary Center of Modeling at the University of Warsaw, Grant g91-1418, g91-1419, g96-1808, g96-1809 and g103-2540 for the availability of high-performance computing resources and support. We acknowledge access to the computing facilities of the Poznan Supercomputing and Networking Center, Grants No. pl0267-01, pl0365-01, pl0471-01 and pl0694-01. A. S. W. acknowledges the support of the National Science Centre, Poland (NCN), through the MINIATURA 9 with project No. 2025/09/X/ST3/00809. J. S. acknowledges funding from the European Union’s Horizon Europe research and innovation programme under the Marie Skłodowska‑Curie grant agreement No 101209345—ART.QM funded by the European Union. Views and opinions expressed are, however, those of the author(s) only and do not necessarily reflect those of the European Union or European Research Executive Agency. Neither the European Union nor the granting authority can be held liable for them. J. M., R. S. and S. S. acknowledge support from the QM4ST project funded by Programme Johannes Amos Commenius, call Excellent Research (Project No. CZ.02.01.01/00/22\_008/0004572). Research at the National Synchrotron Radiation Centre SOLARIS is supported by the Ministry of Science and Higher Education, Poland, under contract no. 1/SOL/2021/2. Measurements were performed at NSRC Solaris under proposal 252163. The research leading to this result has been co-funded by the project NEPHEWS under Grant Agreement No 101131414 from the EU Framework Programme for Research and Innovation Horizon Europe. GS acknowledges support by the Austrian Science Funds (Grant No. AI0656811/21) and the LIT Grant No. LIT-2022-11-SEE-131 of the University of Linz.

\bibliography{references}
\end{document}